\documentclass[fleqn]{article}

\usepackage{amsmath,amssymb}
\usepackage{graphicx}
\usepackage[top=0.75in, bottom=0.75in, left=0.75in, right=0.75in]{geometry}
\usepackage[small]{caption}
\usepackage[noblocks]{authblk}
\usepackage{hyperref}

\usepackage{ulem}
\usepackage{color}

\begin{document}

\title{{\sf Axial Vector $(J^{PC}=1^{++})$ Charmonium and Bottomonium Hybrid Mass Predictions with QCD Sum-Rules}}

\author[1]{D.\ Harnett}
\author[2]{R.T.\ Kleiv}
\author[2]{T.G.\ Steele}
\author[3]{Hong-ying\ Jin}

\affil[1]{Department of Physics, University of the Fraser Valley, Abbotsford, BC, V2S 7M8, Canada}
\affil[2]{Department of Physics and Engineering Physics, University of Saskatchewan, Saskatoon, SK, S7N 5E2, Canada}
\affil[3]{Zhejiang Institute of Modern Physics, Zhejiang University, Zhejiang Province, P. R. China}

\maketitle

\begin{abstract}
Axial vector $(J^{PC}=1^{++})$ charmonium and bottomonium hybrid masses are determined via QCD Laplace sum-rules. Previous sum-rule studies in this channel did not incorporate the dimension-six gluon condensate, which has been shown to be important for $1^{--}$ and $0^{-+}$ heavy quark hybrids. An updated analysis of axial vector charmonium and bottomonium hybrids is presented, including the effects of the dimension-six gluon condensate. The axial vector charmonium and bottomonium hybrid masses are predicted to be 5.13~GeV and 11.32~GeV, respectively. We discuss the implications of this result for the charmonium-like ``XYZ'' states and the charmonium hybrid multiplet structure observed in recent lattice calculations.

\end{abstract}

\section{Introduction}

A long-standing problem in hadron spectroscopy is to determine what role is played, if any, by explicit gluonic constituents in the hadronic spectrum. Quantum chromodynamics (QCD) suggests the possibility of the existence of glueballs which are composed entirely of gluons, as well as hybrids which are composed of a quark, an anti-quark and a gluon. An interesting feature of hybrids is that they can have $J^{PC}$ quantum numbers that are not possible for conventional quark mesons. Consequently, the observation of a state with so-called exotic quantum numbers would be a ``smoking gun'' for the existence of hadrons with explicit gluonic content. Hybrids with non-exotic meson quantum numbers are possible as well; these could signal their presence through supernumerary states in conventional $J^{PC}$ channels. In this work we consider the latter scenario.

Hybrids with non-exotic $J^{PC}$ that contain heavy quarks could coexist with conventional heavy quarkonia states. The large number of anomalous heavy quarkonium-like states discovered above open flavour thresholds has provided an ideal place to look for heavy quark hybrids~\cite{Olsen,Olsen2,Godfrey:2008nc,Pakhlova,Close2007}. A recent review~\cite{Eidelman:2012vu} lists nineteen states discovered since 2003, and many of these states are difficult to accommodate as conventional charmonia~\cite{Barnes2005} leading to numerous suggestions that some of them may be of an exotic nature (see \textit{e.g.} Refs.~\cite{Godfrey:2008nc,Swanson:2006st,Brambilla:2010cs} for reviews).

In this paper we use QCD Laplace sum-rules to investigate axial vector $(J^{PC}=1^{++})$ charmonium and bottomonium hybrids. The constituent gluon model~\cite{Mandula1977} was used for the earliest studies of heavy quark hybrids. Charmonium hybrids have also been studied using the flux tube model~\cite{Barnes1995} which predicts the lightest charmonium hybrids at~4.1--4.2~GeV, as well as lattice QCD~\cite{Perantonis,Liu:2011rn,Liu:2012ze} which gives quenched predictions of about ~4.0~GeV and unquenched predictions of about~4.4~GeV for $1^{++}$ hybrid charmonium. Ref.~\cite{Liu:2012ze} performs a comprehensive study of the charmonium spectrum up to approximately 4.5~GeV and finds evidence for a ground state multiplet of hybrids which contains the $0^{-+}$ and $1^{--}$ states, as well as an excited multiplet containing the $1^{++}$. As far as we are aware, Refs.~\cite{Govaerts:1985fx,Govaerts:1984hc,Govaerts:1986pp} comprise the only QCD sum-rules studies of axial vector charmonium and bottomonium hybrids. 
Several other channels were examined in this work, and many of the resulting sum-rules exhibited instabilities, leading to unreliable mass predictions. The $1^{++}$ channel led to well-behaved sum-rules resulting in mass predictions in the range~4.7--5.7~GeV for hybrid charmonium and~10.9--11.5~GeV for hybrid bottomonium.

In recent sum-rule studies of vector~$(1^{--})$~\cite{Qiao:2010zh} and pseudoscalar~$(0^{-+})$~\cite{Berg:2012gd} heavy quark hybrids it was shown that including the dimension-six gluon condensate can have significant effects on the resulting sum-rules. Specifically, in these channels it was found that inclusion of the dimension-six gluon condensate is sufficient to remove the instabilities observed in Refs.~\cite{Govaerts:1985fx,Govaerts:1984hc,Govaerts:1986pp}. With this paper, we explore the effects of the dimension-six gluon condensate on the sum-rules for axial vector heavy quark hybrids and provide updated mass predictions.

In Section~\ref{theSumRules}, we calculate the appropriate two-point function, including leading-order perturbative contributions and contributions from the dimension-four and dimension-six gluon condensates. In Section~\ref{theAnalysis}, we analyze the sum-rules using the single narrow resonance model and then determine ground state mass predictions. Finally, in Section~\ref{theConclusion}, we discuss the implications of our results for the charmonium-like and bottomonium-like states. With our result for the $1^{++}$ and previous results for the $1^{--}$~\cite{Qiao:2010zh} and $0^{-+}$~\cite{Berg:2012gd} charmonium hybrid mass predictions, we comment on the hybrid multiplet structure identified in Ref.~\cite{Liu:2012ze}.

\section{Laplace Sum-Rules for Axial Vector Heavy Quark Hybrids}
\label{theSumRules} 

The axial vector ($J^{PC}=1^{++}$) heavy quark hybrids may be examined using the following correlation function \cite{Govaerts:1985fx} 
\begin{gather}
\Pi_{\mu\nu}(q)=i\int d^4x \,e^{i q\cdot x}\langle 0\vert T\left[j_\mu(x)j_\nu(0)\right]\vert 0\rangle
\label{basic_corr}
\\
j_\mu=\frac{g}{2}\bar Q\lambda^a\gamma^\nu\tilde G^a_{\mu\nu}Q\,,~\tilde G^a_{\mu\nu}=\frac{1}{2}\epsilon_{\mu\nu\alpha\beta}G^a_{\alpha\beta}\,,
\label{current}
\end{gather} 
with $Q$ representing a heavy quark field. Here we examine the transverse part $\Pi_v$ of \eqref{basic_corr}, which couples to $1^{++}$ states
\begin{equation}
\Pi_{\mu\nu}(q)=\left(\frac{q_\mu q_\nu}{q^2}-g_{\mu\nu} \right)\Pi_v(q^2)+\frac{q_\mu q_\nu}{q^2}\Pi_s(q^2)~.
\label{corr_tensor}
\end{equation}

In Ref.~\cite{Govaerts:1985fx,Govaerts:1984hc} the perturbative and gluon condensate $\langle \alpha G^2\rangle=\langle \alpha G^a_{\mu\nu} G^a_{\mu\nu}\rangle$ contributions to the imaginary part of $\Pi_v$ were calculated to leading order. Here we extend these results by including the contributions of the dimension-six gluon condensate $\langle g^3 G^3\rangle=\langle g^3 f_{abc} G^a_{\mu\nu} G^b_{\nu\alpha} G^c_{\alpha\mu}\rangle$, which were shown to have important consequences for heavy quark hybrid sum-rule studies in different channels~\cite{Qiao:2010zh,Berg:2012gd}. 

First, we verify the leading-order perturbative and $\langle \alpha G^2\rangle$ results \cite{Govaerts:1985fx,Govaerts:1984hc} for $\Pi_v$. We have opted to calculate the full expression for $\Pi_v$ rather than only its imaginary part as in Ref.~\cite{Govaerts:1985fx,Govaerts:1984hc}. This was found to be necessary in order to correctly formulate the sum-rules for the pseudoscalar heavy quark hybrid~\cite{Berg:2012gd}. In addition, verifying existing results using a different approach provides a further consistency check of our results.

\begin{figure}[hbt]
\centering
\includegraphics[scale=0.3]{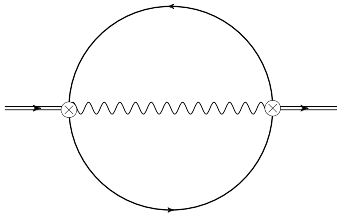}
\caption{Feynman diagram for the leading-order perturbative contribution to $\Pi_v$. The current is represented by the $\otimes$ symbol.
This and all subsequent Feynman diagrams were created with JaxoDraw \cite{Binosi:2003yf}.
}
\label{pert_fig}
\end{figure}

The leading-order perturbative contribution to $\Pi_v$ is represented in Fig.~\ref{pert_fig}. We have made use of the Tarcer \cite{Mertig:1998vk} implementation of loop-integral recurrence relations and tensor structures \cite{Tarasov:1997kx,Tarasov:1996br} to express $\Pi_v$ in terms of a small number of basic integrals. Results for these basic integrals are provided in Refs.~\cite{Boos:1990rg,Davydychev:1990cq,Broadhurst:1993mw}. In $D=4+2\epsilon$ dimensions in the $\overline{{\rm MS}}$ scheme, the perturbative result is
\begin{equation}
\begin{split}
\Pi_v^{{\rm pert}}(q^2)=\frac{m^6\alpha}{8100\pi^3}
&\Biggl[
180(z-1)\left(12z^2-3z+5\right) \phantom{}_3F_2\left(1,1,1;3/2, 3;z\right)
\Biggr.\\
&\Biggl.\qquad
+20z\left(24z^3-96z^2+7z-5\right)
 \phantom{}_3F_2\left(1, 1, 2; 5/2, 4;z\right)
\Biggr] \,, \quad z=\frac{q^2}{4\,m^2} \,,
\end{split}
\label{Pi_pert}
\end{equation}
where terms corresponding to dispersion relation subtraction constants have been omitted. The coupling $\alpha$ and quark mass $m$ implicitly depend on the renormalization scale $\mu$ in the $\overline{{\rm MS}}$ scheme. The generalized hypergeometric functions~\cite{Bateman:1953} in \eqref{Pi_pert} are particularly convenient for sum-rule applications since they clearly reveal the analytic structure of $\Pi_v$, namely a branch cut starting at the threshold $q^2=4m^2$. In addition, the imaginary part may be easily extracted via analytic continuation of the hypergeometric functions. Doing so, we find
\begin{equation}
\begin{split}
{\rm Im}\Pi_v^{\rm pert}(q^2)=
\frac{\alpha m^6}{180\pi^2z^2}
&\Biggl(
\sqrt{z-1} \sqrt{z} \left(15-35z-22z^2-216z^3+48z^4\right)
\Biggr.\\
&\Biggl.\qquad
+15 \left(1-3z+16z^3\right) \log\left[\sqrt{z-1}+\sqrt{z}\,\right]
\Biggr)\,,\quad z>1\,.
\end{split}
\label{Im_Pi_pert}
\end{equation}
We find complete agreement between the integral representation for ${\rm Im}\Pi^{\rm pert}_v$ given in~\cite{Govaerts:1985fx,Govaerts:1984hc} and \eqref{Im_Pi_pert}.

\begin{figure}[hbt]
\centering
\includegraphics[scale=0.3]{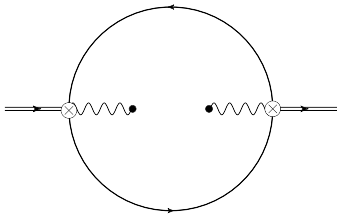}
\caption{Feynman diagram for the leading-order $\langle \alpha G^2\rangle$  contribution to $\Pi_v$.
}
\label{GG_fig}
\end{figure}

The leading-order $\langle \alpha G^2\rangle$ contribution to $\Pi_v$ is represented in Fig.~\ref{GG_fig}. Due to the presence of the field strength in the current \eqref{current} this contribution is most easily calculated using using fixed-point gauge methods (see \textit{e.g.} Ref.~\cite{Elias:1987ac} for examples of this technique). However, plane wave methods could be also be used as they have been proven to be equivalent to fixed-point gauge when gauge-invariant currents such as \eqref{current} are used \cite{Bagan:1992tg}. For the $\langle \alpha G^2\rangle$ contribution we find
\begin{equation}
\Pi^{\rm GG}_v(q^2)=-\frac{ \langle \alpha G^2\rangle}{27\pi} m^2 z(1+2z) \phantom{}_2F_1\left(1, 1; 5/2;z\right)
\,,
\label{Pi_GG}
\end{equation}
where non-physical terms corresponding to dispersion relation subtraction constants have been omitted. 
The imaginary part of \eqref{Pi_GG} is
\begin{equation}
{\rm Im}\Pi^{\rm GG}_v(q^2)=-\frac{ m^2\langle \alpha G^2\rangle}{18}\left(1+2z\right)\frac{\sqrt{z-1}}{\sqrt{z}}\,,\quad z>1
\label{Im_Pi_GG}
\end{equation}
This again agrees with the result given in Refs.~\cite{Govaerts:1985fx,Govaerts:1984hc}.

\begin{figure}[hbt]
\centering
\includegraphics[scale=0.6]{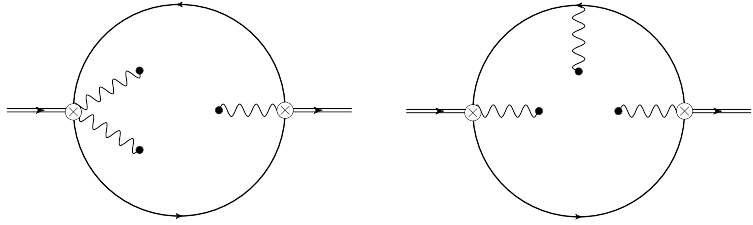}
\caption{Feynman diagram for the leading-order $\langle g^3 G^3\rangle$  contribution to $\Pi_v$. Additional diagrams related by symmetry are not shown.
}
\label{GGG_fig}
\end{figure}

Finally we consider the dimension-six gluon condensate contributions which were not calculated in Refs.~\cite{Govaerts:1985fx,Govaerts:1984hc}. These are represented by the diagrams in Fig.~\ref{GGG_fig}. Again utilizing fixed-point gauge methods, we find
\begin{equation}
\begin{split}
\Pi_v^{\rm GGG}(q^2)=&\frac{\langle g^3G^3\rangle}{1152\pi^2 (z-1)^2}
\left[
2z(2-9z+6z^2)-4z\left(z-1\right)\left(3z-1\right)
\right] \phantom{}_2F_1\left(1, 1; 5/2;z\right)
\\&+
\frac{\langle g^3G^3\rangle}{1152\pi^2 (z-1)^2}
\left[3(17z-9)(z-1)-3(17-46z+27z^2)\right]\,.
\end{split}
\label{Pi_GGG}
\end{equation}
The resulting imaginary part of \eqref{Pi_GGG} is
\begin{equation}
{\rm Im}\Pi_v^{\rm GGG}(q^2)=\frac{\langle g^3G^3\rangle}{384\pi (z-1)^2}\frac{\sqrt{z-1}}{\sqrt{z}}\left[
2(1-3z)(z-1)+(2-9z+6z^2)
\right] \,,\quad z>1\,.
\label{Im_Pi_GGG}
\end{equation} 
The singularity at $z=1$ in \eqref{Im_Pi_GGG} must be dealt with carefully. Although \eqref{Im_Pi_GGG} can be extracted from the purely hypergeometric terms in \eqref{Pi_GGG}, it is not well-defined at $z=1$. This problem is addressed through inclusion of the non-hypergeometric terms in \eqref{Pi_GGG}. These contribute compensating terms which ensure that the contributions of $\Pi^{\rm GGG}(q^2)$ to the sum-rules are well-defined when ${\rm Im}\Pi_v^{\rm GGG}(q^2)$ is integrated from $z=1$. Thus the imaginary part~\eqref{Im_Pi_GGG} by itself is insufficient to construct the contribution of the dimension-six gluon condensate to the QCD Laplace sum-rules.

Now that we have calculated the correlation function, we can proceed with the QCD Laplace sum-rules analysis~\cite{Shifman:1978bx,Shifman:1978by} (for reviews of the methodology see \textit{e.g.} Refs.~\cite{Reinders:1984sr,Narison:2002pw}). Utilizing the standard resonance plus continuum model for the hadronic spectral function, the Laplace sum-rules take the form
\begin{equation}
 {\cal L}_{k}^{\rm QCD}\left(\tau,s_0\right)  = \frac{1}{\pi}\int_{t_0}^{\infty} t^k
   \exp\left[ -t\tau\right] \rho^{\rm had}(t)\; dt \,,
\label{final_laplace}
\end{equation}
where $t_0$ is the hadronic threshold. The quantity on the left hand side of \eqref{final_laplace} is given by
\begin{equation}
{\cal L}_k^{\rm QCD}\left(\tau,s_0\right)\equiv\frac{1}{\tau}\hat B\left[\left(-1\right)^k Q^{2k}\Pi_v\left(Q^2\right)\right] -  \frac{1}{\pi}  \int_{s_0}^{\infty} t^k
   \exp \left[-t\tau  \right]  {\rm Im} \Pi_v(t)\; dt  
\label{laplace}
\end{equation}
where $s_0$ is the continuum threshold, $Q^2=-q^2$ is the Euclidean momentum and $\Pi_v\left(Q^2\right)$ is the axial vector heavy quark hybrid correlation function. The Borel transform operator $\hat B$ is closely related to the inverse Laplace transform~\cite{Bertlmann:1984ih}
\begin{gather}
\frac{1}{\tau}\hat B\left[ f\left(Q^2\right)\right]
=F(\tau)={\cal L}^{-1}
\left[ f\left(Q^2\right)\right] \,,
\\
{\cal L}^{-1}
\left[ f\left(Q^2\right)\right]=\frac{1}{2\pi i}\int\limits_{b-i\infty}^{b+i\infty}
f\left(Q^2\right) e^{Q^2\tau}\,dQ^2
\label{borel_laplace}
\end{gather}
where $b$ is chosen such that $f\left(Q^2\right)$ is analytic to the right of the integration contour in the complex plane.\footnote{Ref.~\cite{Harnett:2000xz} contains detailed examples applying inverse Laplace transform techniques in sum-rule calculations.} Therefore any terms in the full expression for the correlation function $\Pi_v\left(Q^2\right)$ that contribute to the inverse Laplace transform \eqref{borel_laplace} must be included in the construction of the Laplace sum-rules. The singular terms in \eqref{Pi_GGG} that do not contribute to the imaginary part \eqref{Im_Pi_GGG} fall into this category, and thus they are an essential element of the QCD Laplace sum-rules.

Using the results for the leading order perturbative~\eqref{Im_Pi_pert}, $\langle \alpha G^2 \rangle$~\eqref{Im_Pi_GG} and $\langle g^3 G^3 \rangle$~\eqref{Pi_GGG},\eqref{Im_Pi_GGG} contributions, we find
\begin{gather}
\begin{split}
{\cal L}_0^{\rm QCD}\left(\tau,s_0\right)=&\frac{4m^2}{\pi}\int_1^{s_0/4m^2} \left[{\rm Im}\Pi_v^{\rm pert}\left(4m^2 x\right)+{\rm Im}\Pi_s^{\rm GG}\left(4m^2 x\right)\right]\exp{\left(-4m^2\tau x\right)\,dx}
\\
&+\lim_{\eta\to 0^+}\left[\frac{4m^2}{\pi}\int_{1+\eta}^{s_0/4m^2} {\rm Im}\Pi_v^{\rm GGG}(4m^2 x)\exp{\left(-4m^2\tau x\right)\,dx}
+\frac{4m^2\langle g^3G^3\rangle}{192\pi^2\sqrt{\eta}}\exp{(-4m^2\tau)}
\right]
\,,
\end{split}
\label{L_0}
\\
{\cal L}_1^{\rm QCD}\left(\tau,s_0\right)=-\frac{\partial}{\partial\tau}{\cal L}_0^{\rm QCD}\left(\tau,s_0\right)\,.
\label{L_1}
\end{gather}
The terms involving $\eta$ in \eqref{L_1} render the integration in \eqref{L_0} well-defined for the $x\to 1\;\left(\eta\to 0\right)$ limit, and are natural consequence of the inverse Laplace transform of the full expression \eqref{Pi_GGG}. Again, we stress that this expression cannot be obtained with ${\rm Im}\Pi_v^{GGG}$ alone. As before, the mass and coupling in \eqref{L_0} and \eqref{L_1} are functions of the renormalization scale $\mu$ in the $\overline{\rm MS}$-scheme. After evaluating the $\tau$ derivative in \eqref{L_1}, renormalization group improvement may be implemented by setting $\mu=1/\sqrt{\tau}$~\cite{Narison:1981ts}.

\section{Analysis: Mass Predictions for Axial Vector Heavy Quark Hybrids}  
\label{theAnalysis}

In order to extract ground state mass predictions for the $1^{++}$ heavy quark hybrids, we use a single narrow resonance model
\begin{equation}
 \frac{1}{\pi}\rho^{\rm had}(t)=f^2\delta\left(t-M^2\right)\,.
 \label{narrow_res}
\end{equation}
Using this in Eqn.~\eqref{final_laplace} yields
\begin{equation}
{\cal L}_k^{\rm QCD}\left(\tau,s_0\right)=f^2 M^{2k}\exp{\left(-M^2\tau\right)}\,,
\label{narrow_sr}
\end{equation}
from which the ground state mass $M$ can be determined via the ratio
\begin{equation}
M^2=\frac{{\cal L}_1^{\rm QCD}\left(\tau,s_0\right)}{{\cal L}_0^{\rm QCD}\left(\tau,s_0\right)}\,.
\label{ratio}
\end{equation}
It should be noted that the narrow resonance model \eqref{narrow_res} results in a smaller mass prediction $M$ compared to resonance models including width effects~\cite{Elias:1998bq}. Additionally, an upper bound on the ground state mass prediction $M$ can be obtained by taking the limit as $s_0\to\infty$. The resulting upper bound on $M$ is quite robust as it does not depend on the resonance model or how the QCD continuum is modelled.

To extract a ground state mass prediction the sum-rule parameters must be fixed. In the interest of self-consistency, we have chosen to utilize sum-rule estimates of quark masses. For the charm and bottom quark masses we take
\begin{gather}
m_c\left(\mu=m_c\right)=\overline m_c=\left(1.28\pm 0.02\right)\,{\rm GeV}\,,
\label{mc_mass}
\\
m_b\left(\mu=m_b\right)=\overline m_b=\left(4.17\pm 0.02\right)\,{\rm GeV}~,
\label{mb_mass}
\end{gather}
corresponding to the full range of $\overline {\rm MS}$ charm and bottom quark mass estimates of Refs.~\cite{Chetyrkin:2009fv,Narison:2011rn,Narison:2010cg,Kuhn:2007vp} and in agreement with the ranges recommended by the Particle Data Group~\cite{pdg}. We have used one-loop $\overline{\rm MS}$ expressions for the coupling and quark masses. The coupling is evolved from the $\tau$ and $Z$ mass for charmonium and bottomonium hybrids, respectively:
\begin{gather}
\alpha(\mu)=\frac{\alpha\left(M_\tau\right)}{1+\frac{25\alpha\left(M_\tau\right)}{12\pi}\log{\left(\frac{\mu^2}{M_\tau^2}\right)}}
\,,~\alpha\left(M_\tau\right)=0.33\,;
\\
\alpha(\mu)=\frac{\alpha\left(M_Z\right)}{1+\frac{23\alpha\left(M_Z\right)}{12\pi}\log{\left(\frac{\mu^2}{M_Z^2}\right)}}
\,,~\alpha\left(M_Z\right)=0.118\,.
\end{gather} 
The numerical values of $\alpha\left(M_\tau\right)$ and $\alpha\left(M_Z\right)$ are taken from \cite{Bethke:2009jm}, and we use Particle Data Group values of the $\tau$ and $Z$ masses \cite{pdg}. At one-loop order, the $\overline{\rm MS}$ charm and bottom quark masses are given by
\begin{gather}
m_c(\mu)=\overline m_c\left(\frac{\alpha(\mu)}{\alpha\left(\overline m_c\right)}\right)^{12/25}\,,
\\
m_b(\mu)=\overline m_b\left(\frac{\alpha(\mu)}{\alpha\left(\overline m_b\right)}\right)^{12/23}\,.
\end{gather}
For the purposes of the sum-rule analysis we set $\mu=1/\sqrt{\tau}$ as described above. We use the following values of the QCD condensates, extracted from heavy-quark systems \cite{Narison:2010cg}:
\begin{gather}
\langle g^3G^3\rangle=\left(8.2\pm 1.0\right){\rm GeV^2}\langle \alpha G^2\rangle\,
\label{GGG_value}
\\
\langle \alpha G^2\rangle=\left(7.5\pm 2.0\right)\times 10^{-2}\,{\rm GeV^4}\,.
\label{GG_value}
\end{gather}
We find that $\eta=10^{-4}$ is sufficient to numerically evaluate the limit in \eqref{L_0}.

\begin{figure}[hbt]
\centering
\includegraphics[scale=1]{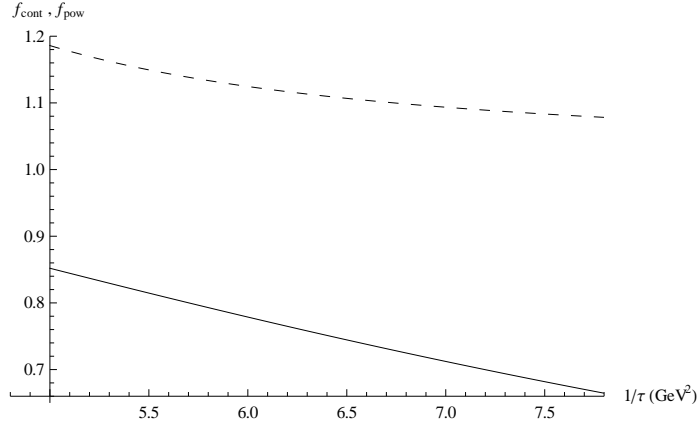}
\caption{The quantities $f_{\text{cont}}\left( \tau,s_0 \right)$ (solid line) and $f_{\text{pow}}\left( \tau,s_0 \right)$ (dashed line) for hybrid charmonium are shown as a function of the Borel scale $1/\tau$ for the optimized value $s_0=33.0\,{\rm GeV^2}$. Central values of the QCD parameters have been used.}
\label{charm_window}
\end{figure}

Now that the numerical values of the physical parameters have been fixed, we may proceed with the sum-rule analysis beginning with hybrid charmonium. First we must establish a region of validity for the sum-rule analysis. To do so, we follow Ref.~\cite{Shifman:1978by} and define the functions
\begin{gather}
f_{\text{cont}}\left( \tau,s_0 \right)=\frac{ {\cal L}_1^{\rm QCD}\left(\tau,s_0\right)/{\cal L}_0^{\rm QCD}\left(\tau,s_0\right) }{{\cal L}_1^{\rm QCD}\left(\tau,\infty\right)/{\cal L}_0^{\rm QCD}\left(\tau,\infty\right) }
\label{f_cont}
\\
f_{\text{pow}}\left( \tau,s_0 \right)=\frac{ {\cal L}_1^{\rm QCD}\left(\tau,s_0\right)/{\cal L}_0^{\rm QCD}\left(\tau,s_0\right) }{{\cal L}_1^{\rm pert}\left(\tau,s_0\right)/{\cal L}_0^{\rm  pert}\left(\tau,s_0\right) }\,,
\label{f_pow}
\end{gather}
where ${\cal L}_k^{\rm pert}$ represents the perturbative contributions to~\eqref{L_0} and~\eqref{L_1}. The functions \eqref{f_cont} and \eqref{f_pow} measure the relative importance of the respective continuum and non-perturbative contributions to the sum-rule. These two functions can be used to constrain the Borel parameter $\tau$ and define a window of reliability for the sum-rule. Inspired by Ref.~\cite{Shifman:1978by}, we impose the constraints $f_{\text{cont}}>0.7$ (\textit{i.e.}, continuum contributions must be less than 30\%) and $|f_{\text{pow}}-1|<0.15$ (\textit{i.e.}, non-perturbative contributions do not exceed 15\%). The purpose of these constraints is to control uncertainties associated with non-perturbative effects (such as truncation of the operator-product expansion at dimension-six) and the continuum model. Figure~\ref{charm_window} depicts the functions $f_{\text{cont}}$ and $f_{\text{pow}}$ for the optimal value of $s_0$ which is determined below. We have also performed the analyis in the 
pole scheme, with a charm quark pole mass of $m_c^{\text{pole}}=1.71\,{\rm GeV}$~\cite{pdg}. As in the pseudoscalar charmonium hybrid analysis~\cite{Berg:2012gd}, we find that the sum-rule window closes rapidly in the pole scheme, and hence the $\overline{{\rm MS}}$ scheme is more suitable for this analysis. The advantage of the $\overline{{\rm MS}}$ scheme is also seen in Ref.~\cite{Jamin:2001fw}.

\begin{figure}[hbt]
\centering
\includegraphics[scale=1]{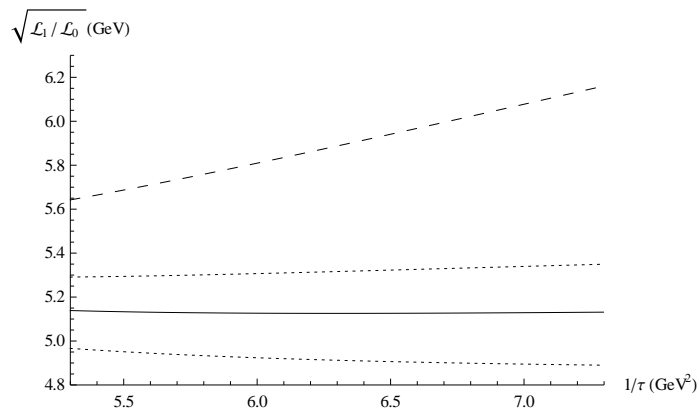}
\caption{The ratio ${\cal L}_1^{\rm QCD}\left(\tau,s_0\right)/{\cal L}_0^{\rm QCD}\left(\tau,s_0\right)$
for hybrid charmonium is shown as a function of the Borel scale $1/\tau$ for the optimized value $s_0=33\,{\rm GeV^2}$ (solid curve). The ratio is also shown for $s_0=38\,{\rm GeV^2}$ (upper dotted curve), $s_0=28\,{\rm GeV^2}$ (lower dotted curve) and $s_0\to\infty$ (uppermost dashed curve). Central values of the QCD parameters have been used.}
\label{charm_opt}
\end{figure}

The optimal $s_0$ value is determined as follows. First, the lowest value of $s_0$ where the mass prediction \eqref{ratio} stabilizes (exhibits a minimum) within its sum-rule window is identified. In this case, we find the minimum value of $s_0$ to be $32\,{\rm GeV^2}$, with a corresponding sum-rule window of $5.3\,{\rm GeV^2} < 1/\tau <7.3\,{\rm GeV^2}$. Figure~\ref{charm_opt} shows the mass prediction \eqref{ratio} within this sum-rule window for several values of $s_0$. Second, we define
\begin{equation}
\chi^2\left(s_0\right)=\sum_j \left( \frac{1}{M}\sqrt{\frac{{\cal L}_1^{\rm QCD}\left(\tau_j,s_0\right)}{{\cal L}_0^{\rm QCD}\left(\tau_j,s_0\right)}}-1\right)^2\,,
\label{chi_sq}
\end{equation}
summed over the window $5.3\,{\rm GeV^2} < 1/\tau <7.3\,{\rm GeV^2}$, and then search for the value of $s_0$ that minimizes \eqref{chi_sq}. The width of the sum-rule window increases slowly as $s_0$ is increased from the minimum value, so this approach guarantees that \eqref{chi_sq} is calculated in a region where the sum-rule is reliable for all values of $s_0$. This procedure results in an optimal $s_0=33\,{\rm GeV^2}$ and a corresponding charmonium hybrid mass prediction of $5.13\,{\rm GeV}$. Note that the limit as $s_0\to\infty$ cannot be used here to obtain a demonstrable upper bound on the charmonium hybrid mass since the mass prediction \eqref{ratio} does not stabilize within the sum-rule window, as can be seen from Fig.~\ref{charm_opt}. On the Borel window, the optimized curve in Fig.~\ref{charm_opt} is virtually $\tau$-independent; this provides us with strong \textit{a posteriori} justification for the use of a single narrow resonance model. Hence any excited states are either exponentially 
suppressed by $\tau$ relative to the ground state or are weakly coupled and absorbed into the continuum.

We now estimate the uncertainty in the charmonium hybrid mass prediction due to uncertainties in the QCD input parameters. Interestingly, the uncertainty in the mass prediction is dominated by $\langle \alpha G^2\rangle$~\eqref{GG_value}, while the uncertainties due to the charm quark mass~\eqref{mc_mass} and $\langle g^3 G^3\rangle$~\eqref{GGG_value} are significantly smaller. This is in contrast with the pseudoscalar heavy quark hybrid, where the uncertainty in the mass prediction is dominated by the dimension-six gluon condensate. We have made no attempt to estimate contributions to these uncertainties from higher loop effects. Since we are interested in hybrids that may exist among the established charmonium-like states, we have explored the effect of resonance widths on our mass predictions with a $200\,{\rm MeV}$ width, corresponding to the widest of these established resonances~\cite{pdg}. Using methods described in Ref.~\cite{Elias:1998bq}, we find that our mass prediction changes by less than $1\% $,
 which is negligible compared to the uncertainty due to the QCD input parameters. Adding the QCD parameter uncertainties in quadrature, we predict the charmonium hybrid mass to be $5.13\pm0.25\,{\rm GeV}$. This prediction is in good agreement with the range of results of 4.7--5.7~GeV found in Refs.\cite{Govaerts:1985fx,Govaerts:1986pp}, all of which were derived from stable sum-rules that did not include effects of the dimension-six gluon condensate. Thus we can conclude that the dimension-six condensate is not as significant in the $1^{++}$ channel as it is in the $0^{-+}$ and $1^{--}$ channels of hybrid charmonium. 

\begin{figure}[hbt]
\centering
\includegraphics[scale=1]{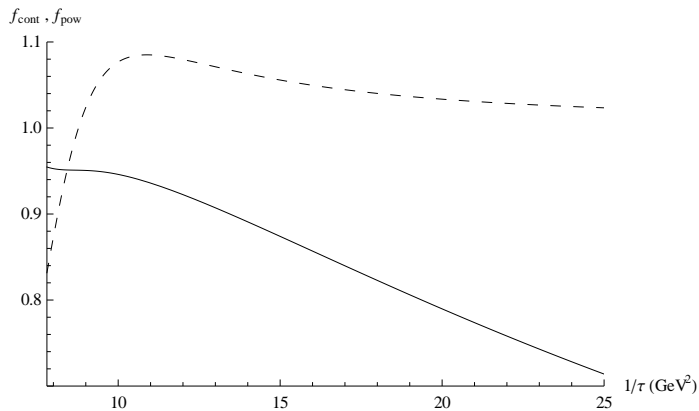}
\caption{The quantities $f_{\text{cont}}\left( \tau,s_0 \right)$ (solid line) and $f_{\text{pow}}\left( \tau,s_0 \right)$ (dashed line) for hybrid bottomonium are shown as a function of the Borel scale $1/\tau$ for the optimized value $s_0=150\,{\rm GeV^2}$. Central values of the QCD parameters have been used.}
\label{bottom_window}
\end{figure}

\begin{figure}[hbt]
\centering
\includegraphics[scale=1]{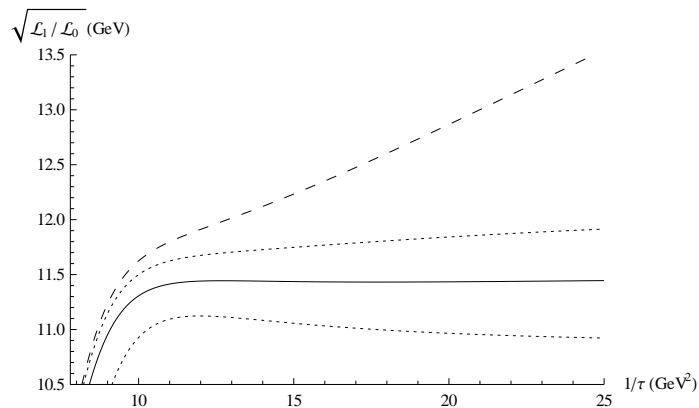}
\caption{The ratio ${\cal L}_1^{\rm QCD}\left(\tau,s_0\right)/{\cal L}_0^{\rm QCD}\left(\tau,s_0\right)$ for hybrid bottomonium is shown  as a function of the Borel scale $1/\tau$ for the optimized value $s_0=150\,{\rm GeV^2}$ (solid curve). For comparison the ratio is also shown for $s_0=170\,{\rm GeV^2}$ (upper dotted curve), $s_0=130\,{\rm GeV^2}$ (lower dotted curve) and  $s_0\to\infty$ (uppermost dashed curve). Central values of the QCD parameters have been used.}
\label{bottom_opt}
\end{figure}

The sum-rule analysis of hybrid bottomonium is very similar. The sum-rule is reliable in the region $s_0 > 145\,{\rm GeV^2}$ and $7.8\,{\rm GeV^2} < 1/\tau <25.0\,{\rm GeV^2}$. The functions $f_{\text{cont}}$~\eqref{f_cont} and $f_{\text{pow}}$~\eqref{f_pow} are shown in this region in Figure~\ref{bottom_window}, and we again use the constraints $0.85<f_{\text{pow}}<1.15$ and $f_{\text{cont}}>0.7$. As in the hybrid charmonium analysis, it is not possible to obtain a demonstrable upper bound on the mass prediction since the ratio~\eqref{ratio} for $s_0\to\infty$ does not stabilize within the sum rule window. Using~\eqref{chi_sq}, we find the optimal $s_0=150\,{\rm GeV^2}$. Figure~\ref{bottom_opt} shows the mass prediction \eqref{ratio} within the sum-rule window for several values of $s_0$. The uncertainty analysis again shows that the error in $\langle \alpha G^2\rangle$ dominates the error in the bottomonium hybrid mass prediction and that resonance width effects are negligible. Adding the errors in quadrature, we predict a bottomonium hybrid mass of $11.32\pm0.32\,{\rm GeV}$. This result is in good agreement with the range of ~10.9--11.5~GeV for hybrid bottomonium predicted in Refs.~\cite{Govaerts:1985fx,Govaerts:1986pp}.

\section{Conclusions}
\label{theConclusion}

In this paper we have studied axial vector ($J^{PC}=1^{++}$) heavy quark hybrids via QCD Laplace sum-rules. We have calculated the full expressions for the leading order perturbative and $\langle \alpha G^2\rangle$ contributions to the correlation function, and have noted that the corresponding imaginary parts of these expressions agree with the results given in Refs.~\cite{Govaerts:1985fx,Govaerts:1986pp}. For the first time, we have also determined the contributions from $\langle g^3G^3\rangle$, which were not included in previous work. For these it was shown that the imaginary part alone was insufficient to formulate the QCD Laplace sum-rules, and the full expression for the $\langle g^3G^3\rangle$ contribution to the correlation function was needed.

In Refs.~\cite{Govaerts:1985fx,Govaerts:1986pp}, many of the sum-rules for various $J^{PC}$ heavy quark hybrids exhibited instabilities, and hence the resulting mass predictions for those channels are unreliable. Recent sum-rule analyses of vector~$(1^{--})$~\cite{Qiao:2010zh} and pseudoscalar~$(0^{-+})$~\cite{Berg:2012gd} heavy quark hybrids have shown that the inclusion of the dimension-six gluon condensate stabilizes the sum-rules for these channels. Although no instabilities were found for $1^{++}$ heavy quark hybrids in Refs.~\cite{Govaerts:1985fx,Govaerts:1986pp}, it is nevertheless of interest to examine the effects of the dimension-six gluon condensate in this channel given the significant effect observed in other channels. Including the $\langle g^3 G^3\rangle$ contributions in our analysis results in the predictions of $5.13\pm0.25\,{\rm GeV}$ for hybrid charmonium and $11.32\pm0.32\,{\rm GeV}$ for hybrid bottomonium. Our results are in agreement with the predictions of Refs.~\cite{Govaerts:1985fx,
Govaerts:1986pp} which ranged from~4.7--5.7~GeV and~10.9--11.5~GeV for axial vector charmonium and bottomonium hybrids, respectively. The uncertainties in the mass predictions are dominated by the uncertainty in $\langle \alpha G^2\rangle$ while the uncertainty due to $\langle g^3G^3\rangle$ is less significant, in contrast to the uncertainty in the pseudoscalar hybrid mass predictions~\cite{Berg:2012gd}. This, together with our agreement with the mass predictions of Refs.~\cite{Govaerts:1985fx,Govaerts:1986pp}, suggests that unlike the vector~\cite{Qiao:2010zh} and pseudoscalar~\cite{Berg:2012gd} channels, the effects of the dimension-six gluon condensate are less significant for the axial vector channel.

To date, all of the charmonium-like ``XYZ'' states have been discovered in the mass range~3.8--4.7~GeV~\cite{Eidelman:2012vu}. Clearly our prediction of $5.13\,\rm{ GeV}$ does not support the identification of any of these states as an axial vector charmonium hybrid. The first discovered charmonium-like state was the X(3872), whose possible $J^{PC}$ assignments are $1^{++}$ or $2^{-+}$~\cite{Abulencia:2006ma,Abe:2005iya}, although the $1^{++}$ option is strongly favoured~\cite{Brambilla:2010cs}. Many different proposals have been made regarding the nature of the X(3872): a conventional charmonium state, a $D^0\,\bar{D}^{0*}$ molecule, a tetraquark and a hybrid (see \textit{e.g.} Ref.~\cite{Swanson:2006st} for a review). The hybrid interpretation was suggested in Ref.~\cite{Li:2004sta}, but has now been largely set aside. The reason for this seems to be that both flux-tube~model~\cite{Barnes1995} and lattice QCD~\cite{Perantonis,Liu:2011rn,Liu:2012ze} predict that the lightest charmonium hybrids have masses 
significantly greater than that of the X(3872). If its quantum numbers are shown to be $1^{++}$, our mass prediction of $5.13\,\rm{ GeV}$ is in agreement with the results of other theoretical approaches that seem to preclude a charmonium hybrid interpretation of the X(3872).

It is interesting to note the large difference between the predicted masses of 3.82~GeV for $0^{-+}$~\cite{Berg:2012gd} and 4.12--4.79 for $1^{--}$~\cite{Qiao:2010zh} hybrid charmonium compared to the $1^{++}$ prediction of 5.13~GeV. In Ref.~\cite{Liu:2012ze} it is suggested that $0^{-+}$ and $1^{--}$ are members of a ground state charmonium hybrid multiplet, while $1^{++}$ is a member of a multiplet of excited charmonium hybrids. Although the mass splittings are significantly larger, the present result and those of Refs.~\cite{Qiao:2010zh,Berg:2012gd} seem to be in approximate agreement with this multiplet structure. Future work to update remaining unstable sum-rule channels in Refs.~\cite{Govaerts:1985fx,Govaerts:1984hc,Govaerts:1986pp} to include the effects of the dimension-six gluon condensate would clarify the predictions for the spectrum of charmonium hybrids from a QCD sum-rules standpoint.

\bigskip
\noindent
{\bf Acknowledgements:}  We are grateful for financial support from the Natural Sciences and Engineering Research Council of Canada (NSERC).


\end{document}